\documentclass[aps,prc,twocolumn,superscriptaddress,showpacs,floatfix]{revtex4}
\usepackage{graphicx,color}
\usepackage{braket}
\usepackage{amsmath}
\usepackage{amssymb}
\usepackage{multirow}
\usepackage{ulem}

\newlength{\figwidth}
\newcommand{\Ve}[1]{\ensuremath{\boldsymbol{#1}}}

\newcommand{\beq}{\begin{equation}}
\newcommand{\eeq}{\end{equation}}
\newcommand{\bea}{\begin{eqnarray}}
\newcommand{\eea}{\end{eqnarray}}
\def\a{\alpha}

\def\la{\mathrel{\mathpalette\fun <}}

\def\fun#1#2{\lower3.6pt\vbox{\baselineskip0pt\lineskip.9pt
\ialign{$\mathsurround=0pt#1\hfil##\hfil$\crcr#2\crcr\sim\crcr}}}

\begin{document}
\setlength{\figwidth}{0.98\columnwidth}

\title{Four-body dynamics in $^6$Li elastic scattering }

\author{Shin Watanabe}
\email[]{s-watanabe@phys.kyushu-u.ac.jp}
\affiliation{Department of Physics, Kyushu University, Fukuoka 812-8581, Japan}

\author{Takuma Matsumoto}
\email[]{matsumoto@phys.kyushu-u.ac.jp}
\affiliation{Department of Physics, Kyushu University, Fukuoka 812-8581, Japan}

\author{Kazuyuki Ogata}
\email[]{kazuyuki@rcnp.osaka-u.ac.jp}
\affiliation{Research Center for Nuclear Physics (RCNP), Osaka University,
Ibaraki 567-0047, Japan}

\author{Masanobu Yahiro}
\email[]{yahiro@phys.kyushu-u.ac.jp}
\affiliation{Department of Physics, Kyushu University, Fukuoka 812-8581, Japan}

\begin{abstract}
We analyze $^6$Li elastic scattering in a wide range 
of incident energies ($E_{\rm in}$), 
assuming the $n+p+\a+{\rm target}$ four-body model and solving 
the  dynamics with the four-body version 
of the continuum-discretized coupled-channels method (CDCC). 
Four-body CDCC  well reproduces the experimental data 
with no adjustable parameter 
for $^{6}$Li+$^{209}$Bi scattering at $E_{\rm in}=24$--50 MeV 
and $^{6}$Li+$^{208}$Pb scattering at $E_{\rm in}=29$--210 MeV. 
In the wide $E_{\rm in}$ range, $^6$Li breakup is significant 
and provides repulsive corrections to the folding potential. 
As an interesting property, $d$ breakup is strongly suppressed 
in $^6$Li-breakup processes independently of $E_{\rm in}$. 
We investigate what causes the $d$-breakup suppression. 
\end{abstract}


\maketitle
\section{Introduction}
\label{Introduction}
Understanding of nucleus-nucleus (AA) scattering is a goal 
in nuclear physics. Recent developments in production 
of radioactive ion beams make this subject more fruitful. 
In particular, elastic scattering is an important part 
of an overall understanding of AA scattering. 
A widespread approach to this subject such as 
distorted-wave Born approximation and 
the continuum discretized coupled-channels method (CDCC) 
\cite{CDCC-review1,CDCC-review2,CDCC-review3} 
is based on the use of an 
optical potential for the description of elastic scattering. 
A pioneering systematic analysis on AA elastic scattering was made by 
Satchler and Love \cite{Satchler-1979,Satchler}. They found that 
AA optical potentials, particularly in its real part, can be described 
by the double-folding model except for weakly-bound projectiles 
such as $^{6}$Li. 
The problem on $^{6}$Li scattering was solved later with CDCC 
by taking account of projectile-breakup effects \cite{CDCC-review1}.

Projectile breakup is essential for scattering of 
weakly-bound nuclei. 
CDCC is a fully quantum-mechanical method for treating dynamics among 
various kinds of channels including breakup (continuum) channels. 
For scattering of deuteron ($d$) on targets (T), 
the effects  are found to be significant 
in a wide range of incident energies, say $E_{\rm in} \la 700$ MeV, by 
assuming the $n+p+\rm{T}$ model and solving 
the three-body dynamics with CDCC \cite{CDCC-review1}. 
Nowadays, three-body dynamics in scattering of two-body projectiles 
is often analyzed by CDCC.

Our interest is now going to four-body dynamics in scattering of 
three-body projectiles. 
CDCC for three- and four-body scattering are now called 
three- and four-body CDCC, respectively. 
Four-body CDCC is an extension of three-body CDCC, 
but the formulation is not straightforward 
since it is not easy to prepare the bound and low-lying 
continuum states of three-body projectile before doing coupled-channel 
calculations. This problem was solved by two approaches; 
one is the combination \cite{Mat10} of the pseudostate discretization 
and the complex scaling method \cite{ABC} 
and the other is the combination \cite{4body-CDCC-bin} of 
the momentum-bin discretization and the hyperspherical harmonics 
method \cite{HRWF}. 
Four-body CDCC is one of state-of-the-art calculations in nuclear physics.

Three-body projectiles have more complicated structure than two-body ones. 
A typical and interesting example is the difference between 
$^6$He and $^6$Li. 
$^6$He is a Borromean nucleus and is well described 
by the $n+n+\alpha$ model. $^6$He has no bound state 
in its two-body subsystems, so that the ground and excited continuum states 
consist of three-body configurations only. 
This property makes four-body dynamics of $^6$He scattering 
relatively simpler. 
Meanwhile, $^6$Li is well described by the $n+p+\alpha$ model and 
has a bound state in the $n+p$ subsystem. 
Therefore, the ground and excited continuum states consist 
of both $d \alpha$ two-body and $np \alpha$ three-body 
configurations.  
In fact, the probability of $d \alpha$ configurations is 
about 70\% in the ground state. 
This situation makes it more difficult to understand four-body dynamics 
of $^6$Li scattering. 
Four-body CDCC was first applied to a simpler case, i.e., $^6$He scattering. 
The analysis was successful in reproducing the experimental data 
with no adjustable parameter for both elastic and breakup cross sections 
\cite{Mor01,Mat03,Mat04,Ega04,Rod05,Mor06,Mat06,Rod08,Mor09,4body-CDCC-bin,Ega09,Mat09,Mat10}.

$^6$Li-breakup effects were first analyzed 
with three-body CDCC based on the $d+\alpha+{\rm T}$ model 
\cite{CDCC-review1}. 
This analysis showed that $^6$Li breakup effects provide large repulsive 
corrections to the folding potential. 
This is the reason why 
$^6$Li elastic scattering was not described by the double-folding 
model. However, this statement should be reinforced by 
four-body CDCC.

$^6$Li~+~$^{209}$Bi elastic scattering at $E_{\rm in}=29.9$ and 32.8 MeV near 
the Coulomb-barrier energy $E_{\rm b}^{\rm Coul} \approx 30$ MeV 
were first analyzed with three-body CDCC \cite{Kee03}; 
note that $E_{\rm in}$ stands for an incident energy in the laboratory system. 
However, the three-body CDCC calculation could not account 
for measured elastic cross sections without introducing a normalization factor 
0.8 to $d$-T and $\alpha$-T optical potentials. 
The problem was solved by four-body CDCC based on the 
$n+p+\a+{\rm T}$ model \cite{Wat12}. 
In fact, the calculation reproduced the experimental 
data with no adjustable parameter. As an interesting result, it was reported 
that $d$ breakup is strongly suppressed in $^6$Li breakup processes of 
the elastic scattering. 
The failure of three-body CDCC comes from the use of the 
phenomenological $d$-T optical potential 
that includes $d$-breakup effects implicitly. 
In fact, if the $d$-$\a$ potential is replaced by the single folding potential 
obtained by folding $p$-T and $n$-T optical potentials 
with the $d$ ground state, the three-body CDCC calculation well reproduces 
the experimental data. 
Thus, the $d$-breakup suppression is a key to 
understanding four-body dynamics in $^6$Li scattering. 
The next question to be addressed is whether the $d$-breakup suppression is 
realized also at $E_{\rm in} > E_{\rm b}^{\rm Coul}$, 
even if $d$ breakup is important for 
the corresponding $d$ scattering at the same incident
 energy per nucleon (the same incident velocity) \cite{CDCC-review1}.

In this work, four-body ($n+p+\a+{\rm target}$) dynamics 
of $^6$Li elastic scattering is analyzed over a wide range of $E_{\rm in}$ 
with four-body CDCC. 
Four-body CDCC reproduces the experimental data without introducing 
any adjustable parameter for $^{6}$Li+$^{209}$Bi scattering 
at $E_{\rm in}=24$--50 MeV and $^{6}$Li+$^{208}$Pb scattering 
at $E_{\rm in}=29$--210 MeV. 
We can then investigate the four-body dynamics clearly. 
In the present $E_{\rm in}$ range, $^6$Li breakup is significant and 
provides repulsive corrections to the folding model. 
The $d$-breakup suppression is always realized in the $E_{\rm in}$ range. 
We then investigate what causes the $d$-breakup suppression.

In the present work, the theoretical framework consists of four-body CDCC 
for reaction calculations and the Gaussian expansion method (GEM) \cite{Hiy03} 
for structure calculations. This framework is recapitulated in 
Sec. \ref{Theoretical framework}.  
In Sec. \ref{Results}, we present the results of four-body CDCC calculations 
and discuss the nature of the $d$-breakup suppression. 
Section \ref{summary} is devoted to a summary.

\section{Theoretical framework}
\label{Theoretical framework}

\subsection{Four-body CDCC}
\label{Overview}
We recapitulate four-body CDCC for $^6$Li scattering 
from a target nucleus (T); 
see Ref.~\cite{CDCC-review3} for the detail. 
Since $^6$Li is well described by the $n+p+\alpha$ three-body 
model, we consider the $n+p+\alpha+\mathrm{T}$ four-body system for 
$^6$Li scattering. 
The scattering state $\Psi$ with the total energy $E$ is then 
governed by the four-body Schr\"{o}dinger equation 
\begin{equation}
(H_4-E)\Psi=0,\label{eq:4b-Schroedinger}
\end{equation}
with the total Hamiltonian 
\begin{equation}
H_4=K_{\Ve R}+U_n+U_p+U_\alpha+\frac{e^2Z_{\rm Li}Z_{\rm T}}{R}+h_{np\alpha},\label{eq:H4-ver2}
\end{equation}
where $h_{np\alpha}$ denotes the internal Hamiltonian of $^6$Li, 
 $K_{\Ve R}$ stands for the kinetic energy operator 
with respect to the relative coordinate ${\Ve R}$ between $^6$Li and T,
 and $U_{x}$ ($x=n,p,\alpha$) represents 
the optical potential between $x$ and T. 
In Eq. (\ref{eq:H4-ver2}), the Coulomb breakup is neglected
and the Coulomb interactions of $p$-T and $\alpha$-T are then approximated into 
$e^2Z_{\rm Li}Z_{\rm T}/R$, 
where $Z_{\rm A}$ is the atomic number of nucleus A.
This approximation is performed in all the calculations except for Fig.~\ref{fig:el_Coul},
and its accuracy is discussed in Appendix~\ref{sec:CoulombBU} and \ref{sec:E1}.

In CDCC, Eq.~(\ref{eq:4b-Schroedinger}) is solved 
in the model space $P$ spanned by the ground and discretized continuum states  
that are obtained by diagonalizing $h_{np\alpha}$ with $L^2$-type basis functions: 
\begin{equation}
P=\sum_{\gamma=0}^{N}\ket{\Phi_\gamma}\bra{\Phi_\gamma},
\end{equation}
where $\Phi_\gamma$ represents the $\gamma$-th eigenstate with an 
eigenenergy $\varepsilon_\gamma$, i.e., $\Phi_0$ is 
the ground state of $^{6}$Li and the 
$\Phi_\gamma$ for $\gamma=1$--$N$ mean discretized continuum states 
of $^{6}$Li. 
This model-space assumption reduces Eq.~(\ref{eq:4b-Schroedinger}) to 
\begin{equation}
P(H_4-E)P\Psi_\mathrm{CDCC}=0,\label{eq:4b-CDCC}
\end{equation}
for the CDCC wave function
\begin{equation}
\Psi_\mathrm{CDCC}=\sum_{\gamma=0}^{N}\chi_\gamma(\Ve R)\ket{\Phi_\gamma},\label{eq:CDCC-wf}
\end{equation}
where the expansion coefficient $\chi_\gamma$ describes the relative motion 
between T and $^6$Li in its $\gamma$-th state. 
Equation \eqref{eq:4b-CDCC} leads to a set of coupled equations for 
$\chi_\gamma$: 
\begin{equation}
[K_{\Ve R}+U_{\gamma\gamma}-(E-\varepsilon_\gamma)]\chi_\gamma(\Ve R)
=-U_{\gamma\gamma'}\chi_{\gamma'}(\Ve R),
\end{equation}
with the coupling potentials 
\begin{equation}
U_{\gamma\gamma'}=\braket{\Phi_\gamma|U_n+U_p+U_\alpha|\Phi_{\gamma'}}
+\frac{e^2Z_{\rm Li}Z_{\rm T}}{R}\delta_{\gamma\gamma'}.
\end{equation}
This CDCC equation is solved under the standard boundary condition.

\subsection{Structure of $^6$Li in GEM}
\label{GEM}

\begin{figure}[htbp]
\includegraphics[width=0.9\figwidth,clip]{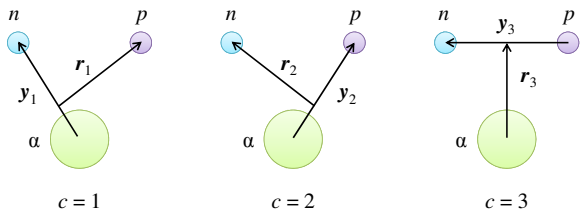}
\caption{(Color online) Three sets of Jacobi coordinates
${\Ve \xi}_c=\{{\Ve r}_c,{\Ve y}_c\}$ 
in the $n$ + $p$ + $\alpha$ three-body system. 
Each set is identified by a label $c$. 
\label{fig:6Li-coordinate}}
\end{figure}

We construct the $\Phi_\gamma$ by applying 
the Gaussian expansion method (GEM) \cite{Hiy03} to the $n+p+\alpha$ system.
The calculation procedure for $^6$Li is the same as 
that for $^6$He in Ref.~\cite{Mat04,Mat06}, although the spin-parity of 
$\Phi_0$ is $1^+$ for $^6$Li but $0^+$ for $^6$He. 
In the GEM, three kinds of Jacobi coordinates, 
 ${\Ve \xi}_c=\{{\Ve r}_c,{\Ve y}_c\}$ for $c=1$--3, are taken as shown 
in Fig.~\ref{fig:6Li-coordinate}. 
 Thanks to this model setting, 
 $^5$He-$p$, $^5$Li-$n$, $d$-$\alpha$, and $n$-$p$-$\alpha$ configurations
 are well incorporated, and thereby fast convergence is obtained 
with respect to expanding the model space $P$.

The model Hamiltonian $h_{np\alpha}$ is defined by 
\begin{align}
h_{np\alpha}&=T_{{\Ve r}_c}+T_{{\Ve y}_c}+V_{n\alpha}+V_{p\alpha}+V_{np}+V_\mathrm{PF} 
\end{align}
with the kinetic energy operator $T_{\Ve x}$ of coordinate ${\Ve x}={\Ve r}_c, 
{\Ve y}_c$ and the interaction $V_\mathrm{ab}$ between particles a and b. 
We take the so-called KKNN interaction~\cite{Kan79} 
for $V_{n\alpha}$ and $V_{p\alpha}$ and the Bonn-A interaction \cite{Mac89} 
for $V_{np}$. 
The interactions well reproduce the corresponding low-energy scattering data.
The operator 
\begin{align}
V_\mathrm{PF}&=\lim_{\lambda_\mathrm{PF} \to \infty}\lambda_\mathrm{PF}
\sum_{c=1}^2\ket{\phi_\mathrm{PF}({\Ve y}_c)}\bra{\phi_\mathrm{PF}({\Ve y}_c)} 
\end{align}
is introduced to exclude the Pauli forbidden states $\phi_\mathrm{PF}$, 
where $\lambda_\mathrm{PF}=10^6$ MeV is taken in actual calculations.

In the GEM, the eigenstates are described as 
\begin{equation}
\Phi_{\gamma} \equiv \Phi_{nI^\pi}({\Ve \xi})=\sum_{c=1}^3\psi_{nI^\pi}^{(c)}({\Ve y}_c,{\Ve r}_c) \label{eq:6Li-wf}
\end{equation}
with the Gaussian basis functions 
\begin{equation}
\psi_{nI^\pi}^{(c)}({\Ve y}_c,{\Ve r}_c)
=\sum_{\lambda\ell\Lambda S}
\sum_{i=1}^{i_\mathrm{max}}\sum_{j=1}^{j_\mathrm{max}}A_{i\lambda j\ell\Lambda S}^{(c)nI^\pi}
\varphi_{i\lambda j\ell\Lambda S}^{(c)}({\Ve y}_c,{\Ve r}_c) , 
\end{equation}
\begin{align}
\varphi_{i\lambda j\ell\Lambda S}^{(c)}({\Ve y}_c,{\Ve r}_c)
&=
y_c^\lambda e^{-(y_c/\bar{y}_i)^2}r_c^\ell e^{-(r_c/\bar{r}_j)^2}\nonumber\\
&\hspace{-20mm}\times\Bigr[
\bigl[Y_{\lambda}(\Omega_{y_c})\otimes Y_{\ell}(\Omega_{r_c})\bigr]_\Lambda
\otimes\bigr[\eta_{1/2}^{(n)}\otimes \eta_{1/2}^{(p)}\bigr]_S\Bigl]_{I^\pi},
\end{align}
where the index $i$ ($j$) means the $i$-th ($j$-th) basis function for the coordinate $y$ ($r$), the symbol 
$\lambda$ ($\ell$) denotes the angular momentum regarding ${\Ve y}$ 
(${\Ve r}$), and $\Lambda$ stands for the total angular momentum.
Furthermore, $\eta_{1/2}^{(n)}$ and $\eta_{1/2}^{(p)}$ denote 
the spin wave functions for $n$ and $p$, and the total spin $S$ of 
the $p+n$ system is set to 1. 
In actual calculations, $\lambda$ and $\ell$ are truncated 
at $\lambda=\lambda_\mathrm{max}$ and $\ell=\ell_\mathrm{max}$, respectively.
The range parameters of Gaussian basis functions are taken 
in the geometric progression: 
\begin{align}
\bar{y}_i&=\bar{y}_1(\bar{y}_\mathrm{max}/\bar{y}_1)^{(i-1)/(i_\mathrm{max}-1)},\\
\bar{r}_j&=\bar{r}_1(\bar{r}_\mathrm{max}/\bar{r}_1)^{(j-1)/(j_\mathrm{max}-1)} \end{align}
with $i_\mathrm{max}=j_\mathrm{max}=10$. 
The range parameters ($\bar{y}_1$, $\bar{y}_\mathrm{max}$, $\bar{r}_1$, $\bar{r}_\mathrm{max}$)  are shown in Table~\ref{tb:maxval}, together with 
the values of $\lambda_\mathrm{max}$ and $\ell_\mathrm{max}$.

\begin{table}[htbp]
\begin{tabular}{cccccccc}
\hline \hline
\makebox[25pt][c]{$c$} & \makebox[25pt][c]{$I^\pi$} & \makebox[25pt][c]{$\lambda_\mathrm{max}$}
 & \makebox[25pt][c]{$\ell_\mathrm{max}$} & \makebox[25pt][c]{$\bar{y}_1$}
 & \makebox[25pt][c]{$\bar{y}_\mathrm{max}$} & \makebox[25pt][c]{$\bar{r}_1$}
 & \makebox[25pt][c]{$\bar{r}_\mathrm{max}$} \\
   &       &   &   &(fm) & (fm) &(fm) & (fm) \\
\hline
 3   & $1^+$ & 2 & 2 & 0.1 & 12.0 & 0.5 & 12.0 \\
 1,2 & $1^+$ & 2 & 2 & 0.5 & 12.0 & 0.5 & 12.0 \\
 3   & $2^+$ & 3 & 2 & 0.1 & 12.0 & 0.5 & 12.0 \\
 1,2 & $2^+$ & 2 & 2 & 0.5 & 12.0 & 0.5 & 12.0 \\
 3   & $3^+$ & 4 & 2 & 0.1 & 12.0 & 0.5 & 12.0 \\
 3   & $3^+$ & 4 & 2 & 0.5 & 12.0 & 0.5 & 12.0 \\
\hline \hline
\end{tabular}
\caption{Model space of GEM calculations. 
The maximum angular momenta 
and the Gaussian range parameters are presented 
for each Jacobi coordinate.
}
\label{tb:maxval}
\end{table}

The effective three-body force 
\begin{equation}
V^\mathrm{3body}(y_1,y_2)=V_3e^{-\nu(y_1^2+y_2^2)}
\end{equation} 
is added to $h_{np\alpha}$ so that the theoretical results can reproduce 
measured binding energy 
($\varepsilon_0$)~\cite{Til00} and 
root mean square radius ($R_\mathrm{rms}$)~\cite{Dob00} of $^{6}$Li; 
in the present case, 
the optimal parameter set is ${V_3=-1.05}$~MeV and ${\nu=0.01}$~fm${}^{-2}$. 
Diagonalizing $h_{np\alpha}+V^\mathrm{3body}$ 
with the Gaussian basis functions, we obtain the $\Phi_\gamma$. 
The theoretical results are summarized in Table~\ref{tb:gs}.

\begin{table}[htbp]
\begin{tabular}{c|ccc}
\hline \hline
\makebox[45pt][c]{} & \makebox[45pt][c]{$I^\pi$} 
& \makebox[45pt][c]{$\varepsilon_0$ (MeV)} & \makebox[45pt][c]{$R_\mathrm{rms}$ (fm)}\\
\hline
 Calc. & $1^+$   & $-3.69$   & 2.43 \\
 Exp.  & $1^+$   & $-3.6989$ & 2.44$\pm$0.07 \\ 
\hline \hline
\end{tabular}
\caption{The spin-parity $(I^\pi)$, the energy $(\varepsilon_0)$
and the matter root mean square radius $(R_\mathrm{rms})$ 
of the $^6$Li ground state calculated with the GEM. 
The experimental data are taken from Refs.~\cite{Til00,Dob00}.
}
\label{tb:gs}
\end{table}

\begin{figure}[htbp]
\includegraphics[width=0.9\figwidth,clip]{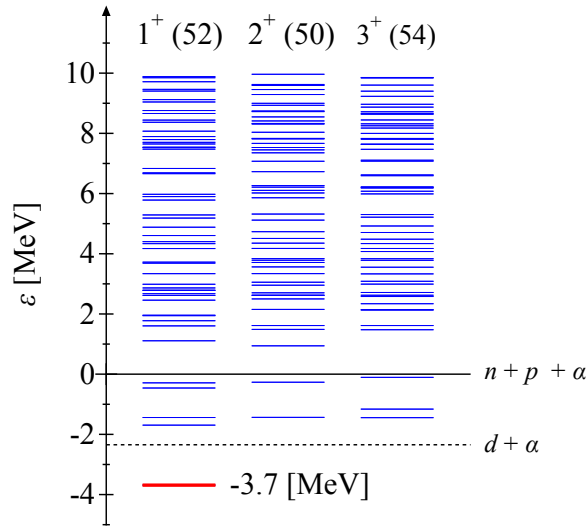}
\caption{(Color online) Calculated eigenenergies of $^6$Li
for $I^\pi=1^+$, 2$^+$, and 3$^+$ from 
the $np\alpha$ threshold ($\varepsilon_\mathrm{th}^{(np\alpha)}=0$ MeV). 
The $d\alpha$ threshold 
($\varepsilon_\mathrm{th}^{(d\alpha)}=-2.2$ MeV)
 is also shown by the dotted line.
The numbers of pseudostates up to $\varepsilon_\mathrm{max}=10$ MeV
 are shown in the parentheses.
\label{fig:state}}
\end{figure}

The resultant eigenenergies are illustrated in Fig.~\ref{fig:state}, 
together with the $d+\alpha$ two-body threshold energy 
$\varepsilon_\mathrm{th}^{(d\alpha)}=-2.2$ MeV and 
the $n+p+\alpha$ three-body one $\varepsilon_\mathrm{th}^{(np\alpha)}=0$ MeV. 
The ground states and the discretized breakup states of $1^+$, $2^+$, $3^+$ 
with $\varepsilon < \varepsilon_\mathrm{max}=10$ MeV are taken 
as the model space $P$ in CDCC calculations. 
We confirmed that this model space yields good convergence 
for the present $^6$Li elastic scattering.

\section{Results}
\label{Results}

\subsection{$^6$Li + $^{209}$Bi elastic scattering}
\label{6Li-209Bi-elastic}

 First, $^6$Li+$^{209}$Bi elastic scattering is 
 analyzed at $E_{\rm in}=24$--50 MeV  with four-body CDCC. 
 As for $U_n$, we take the potential of Koning and Delaroche~\cite{Kon03}, 
 but the spin-orbit interaction is neglected for simplicity. 
 The central potential almost reproduces measured differential cross sections 
 of $n$+$^{209}$Bi scattering at 5 MeV, but the agreement is not perfect. 
 We then made a fine tuning by slightly modifying the parameter 
set~\cite{Wat12}; the resulting parameters are $a_V=0.55$ fm, $W_V=0$ MeV, 
and $W_D=4.0$ MeV. For simplicity, the same parameter set is taken 
for all the incident energies, and $U_p$ is assumed to have the same geometry 
as $U_n$. 
The potential $U_\alpha$ is taken from Ref.~\cite{Bar74} determined 
from measured differential cross sections of $^4$He + $^{209}$Bi scattering 
at 19--22 MeV.

\begin{figure}[htbp]
\includegraphics[width=0.8\figwidth,clip]{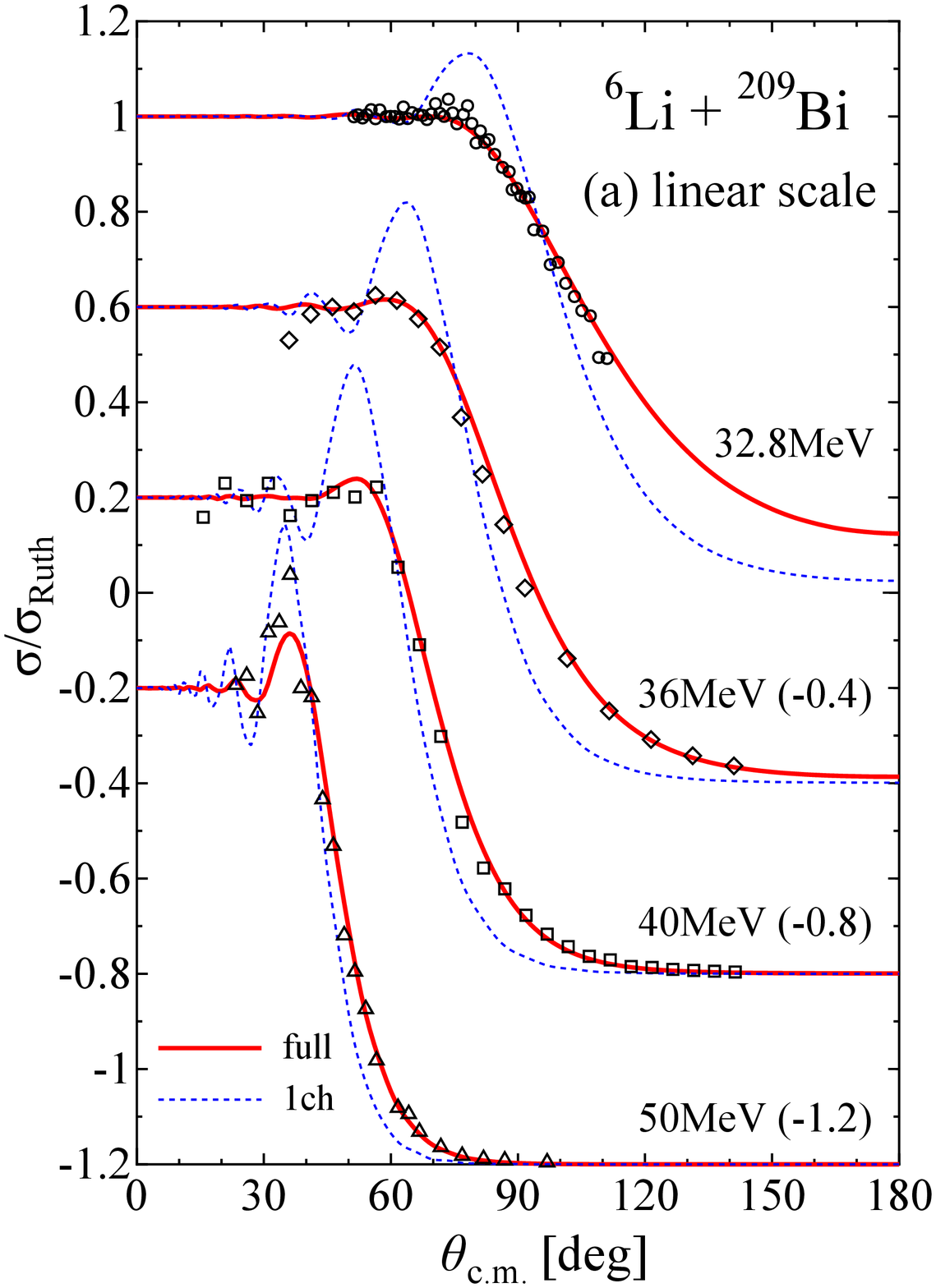}
\includegraphics[width=0.8\figwidth,clip]{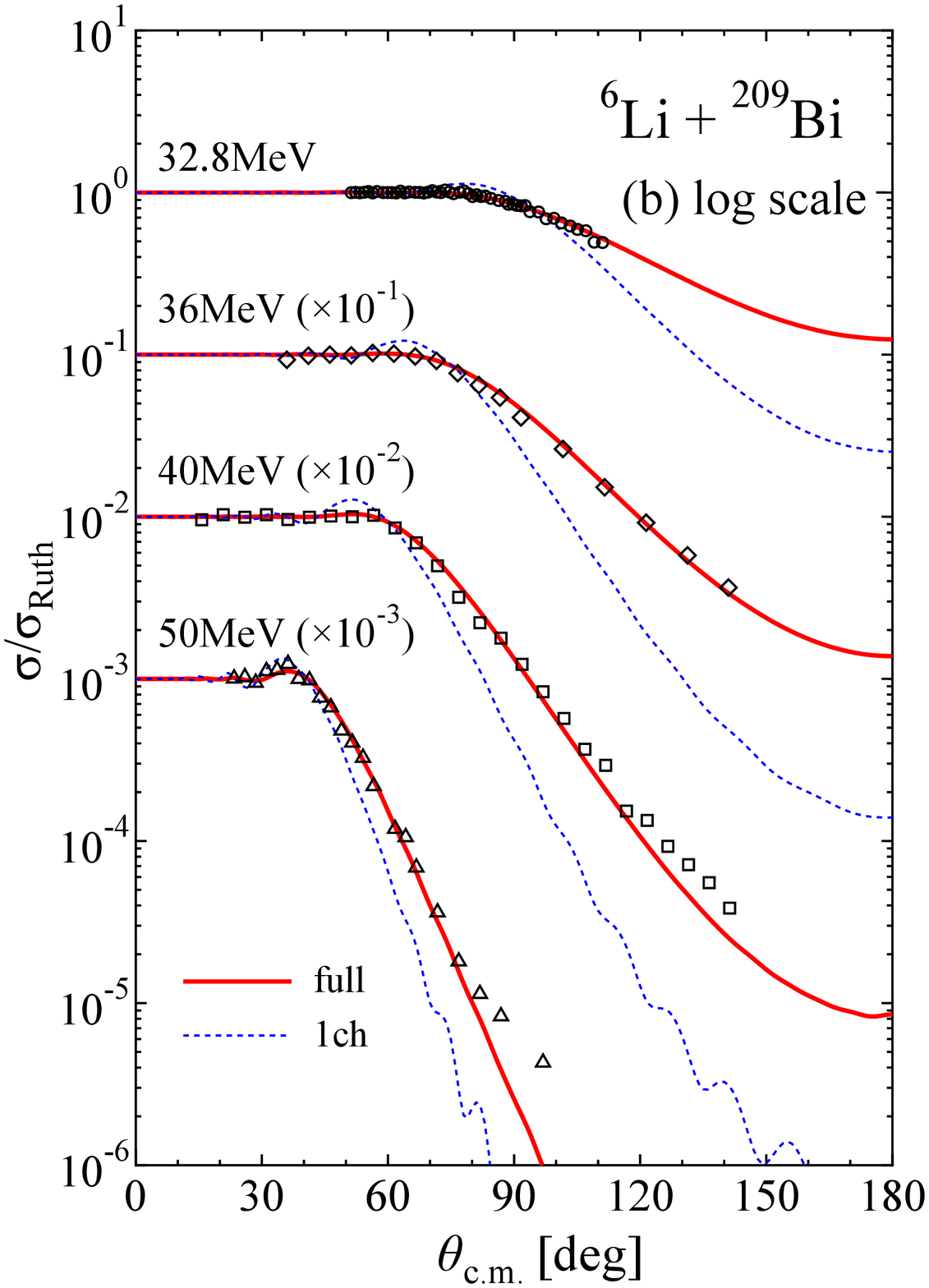}
\caption{(Color online) Elastic cross sections (normalized by the Rutherford cross section) for $^6$Li + $^{209}$Bi scattering at 32.8--50 MeV 
in (a) the linear scale and (b) the logarithmic scale.
 The solid lines represent the results of full four-body CDCC calculations, 
 whereas the dotted lines show the results of 1ch calculations with no breakup effect.
 The symbols are experimental data taken from Refs.~\cite{Agu00,Agu01,San11}.
\label{fig:el}}
\end{figure}

Differential cross sections are plotted as a function of scattering 
angle $\theta_{\rm cm}$ in Fig.~\ref{fig:el} 
for $^6$Li + $^{209}$Bi scattering at $E_{\rm in}=32.8$--50 MeV that is 
larger than the Coulomb-barrier energy $E_{\rm b}^{\rm Coul} \approx 30$ MeV. 
The experimental data are taken from Refs.~\cite{Agu00,Agu01,San11}. 
Four-body CDCC calculations (solid lines) reproduce the data 
with no adjustable parameter. 
The dashed lines denote the results of one-channel (1ch) calculations with 
no breakup effect. The difference between the solid and dashed lines is large, 
indicating that $^{6}$Li breakup, i.e., four-body dynamics is important 
at $E_{\rm in}>V_\mathrm{b}$. 
The scattering angle $\theta_r$ 
at which the dashed line becomes maximum 
nearly corresponds to a rainbow angle in a semi-classical picture, 
and the scattering angle $\theta_g$ at which the Rutherford ratio is 1/4 
approximately corresponds to a grazing angle. 
$^{6}$Li-breakup effects  suppress a strong diffraction pattern 
of the dashed line at $\theta_{\rm cm}$ around $\theta_r$ and 
enhances the cross section largely at $\theta_{\rm cm}$ around $\theta_g$; 
see Fig.~\ref{fig:el}(a) for the suppression and Fig.~\ref{fig:el}(b) 
for the enhancement.

\begin{figure}[htbp]
\includegraphics[width=0.8\figwidth,clip]{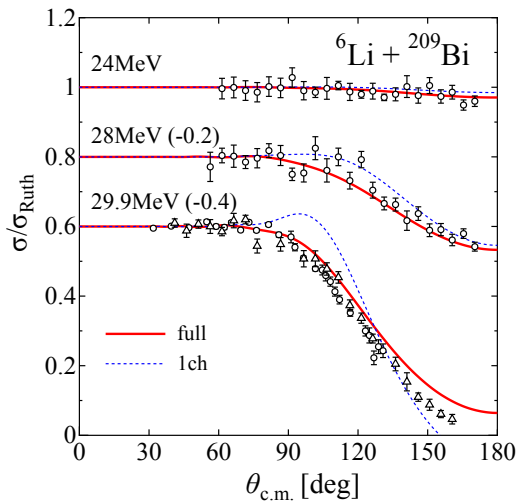}
\caption{(Color online) Same as Fig.~\ref{fig:el}, but 
$E_{\rm in}=24, 28$ and 29.9 MeV. 
\label{fig:el_low}}
\end{figure}

\begin{figure}[htbp]
\includegraphics[width=0.8\figwidth,clip]{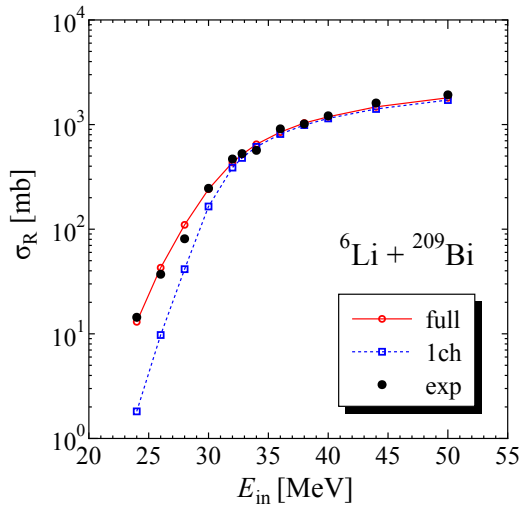}
\caption{(Color online) Total reaction cross section 
$\sigma_{\rm R}$ as a function of $E_\mathrm{in}$.
The solid and dotted line represents the calculation with and without
the channel coupling effects. The derived cross section based on the optical model
analysis~\cite{San11} is also shown.
\label{fig:RCS_log}}
\end{figure}

Figure~\ref{fig:el_low} shows the same figure as Fig.~\ref{fig:el}, 
but $E_{\rm in}$ is smaller than $V_\mathrm{b}$. Again, four-body CDCC 
calculations well account for measured differential cross sections. 
$^{6}$Li breakup effects become small as $E_{\rm in}$ decreases from $V_\mathrm{b}$. In contrast, for the total reaction cross section $\sigma_{\rm R}$, 
the effects are more significant as $E_{\rm in}$ 
goes down from $V_\mathrm{b}$, 
as shown in Fig.~\ref{fig:RCS_log}. 
Four-body dynamics is thus essential for both $E_{\rm in} < V_\mathrm{b}$ and 
$E_{\rm in} > V_\mathrm{b}$.

$^6$Li + $^{209}$Bi scattering at $E_{\rm in}=29.9$ and 32.8 MeV were 
already analyzed by four-body CDCC in our previous work \cite{Wat12}. 
The present calculations are improved from the previous one 
at the following two points. 
In the previous work, the matter radius $R_{\rm rms}$ of $^{6}$Li was 
2.34 fm and slightly underestimates the experimental data 
$R_{\rm rms}=2.44 \pm 0.07$ fm. This problem is solved  
by taking a different parameter set for $V^{\rm 3body}$; 
in the present case $R_{\rm rms}=2.43$ fm, as shown in Table \ref{tb:gs}. 
The second point is the accuracy of numerical calculations. 
In the previous work, the diagonal potentials $U_{\gamma \gamma}$ between 
$2^{+}$ breakup states were numerically inaccurate. This problem is also 
solved. These two corrections are small, so that the present results are 
very close to the previous ones.

\subsection{$^6$Li + $^{208}$Pb elastic scattering}
\label{6Li-208Pb-elastic}

We make the same analysis for a $^{208}$Pb target, 
since the experimental data are available up to the high incident energy
of $E_{\rm in}=210$ MeV.
 Again, we take the central part of 
the Koning-Delaroche potential for $U_n$, and as $U_{\a}$ we use 
the optical potentials of Ref.~\cite{Bar74} for $E_{\rm in}=29$ and 39 MeV 
and those of Ref.~\cite{Per76} for $E_{\rm in}=73.7$ and 210 MeV; 
note that $U_p$ has the same geometry as $U_n$.

\begin{figure}[htbp]
\includegraphics[width=0.8\figwidth,clip]{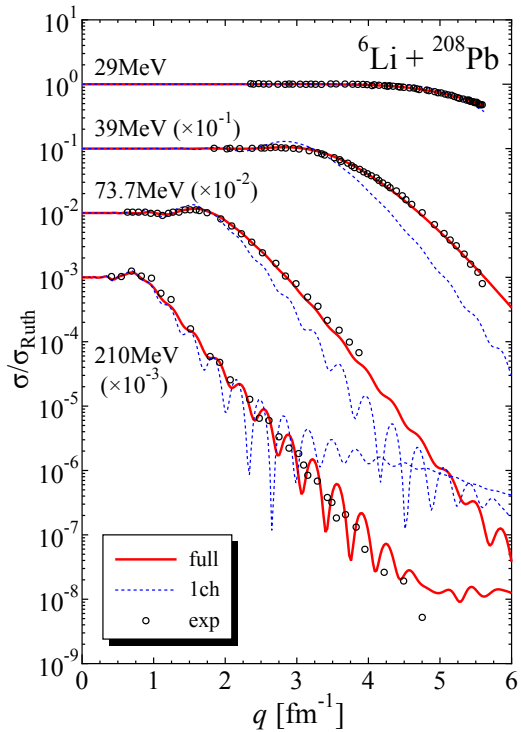}
\caption{(Color online) Elastic cross sections (normalized by the Rutherford cross section)
 for $^6$Li + $^{208}$Pb scattering at 29--210 MeV as a function of the transfer momentum $q$.
 The solid lines represent the results of full four-body CDCC 
 calculations, whereas the dotted lines denote the results of 1ch calculations. 
 The experimental data are taken from Refs.~\cite{Kee94,Huf80,Nad89}.
\label{fig:el_208Pb_log_q}}
\end{figure}

Figure~\ref{fig:el_208Pb_log_q} shows the angular distribution
 of elastic cross sections for $^6$Li + $^{208}$Pb
 scattering at $E_{\rm in}=29$, 39, 73.7 and 210 MeV. 
The experimental data are taken from Refs~\cite{Kee94,Huf80,Nad89}.
This scattering are also well explained by four-body CDCC over a wide range 
of $E_{\rm in}=29$--210 MeV in virtue of projectile-breakup effects. 
 
\begin{figure}[htbp]
\begin{center}
 \includegraphics[width=0.4\textwidth,clip]{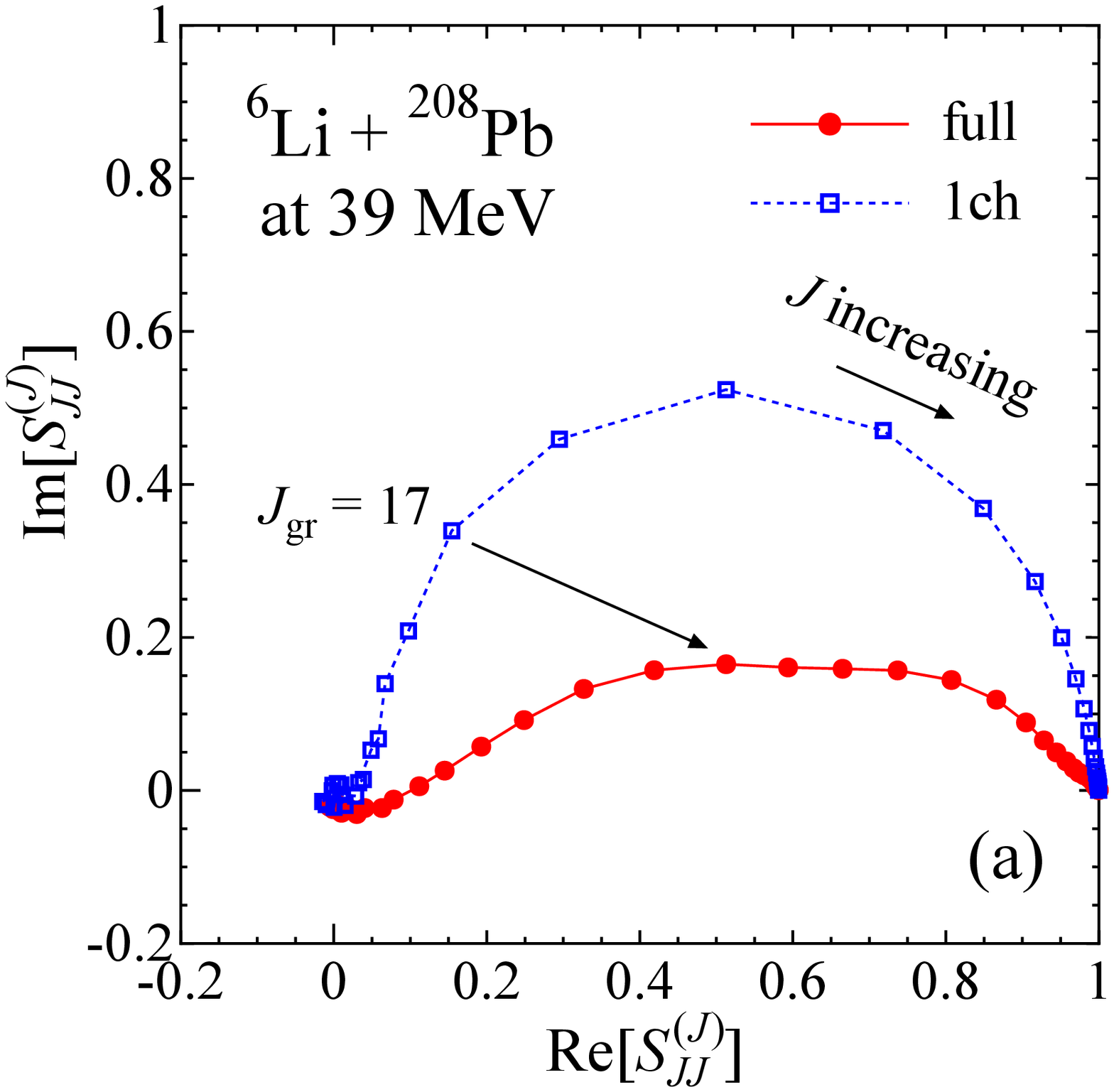}
 \includegraphics[width=0.4\textwidth,clip]{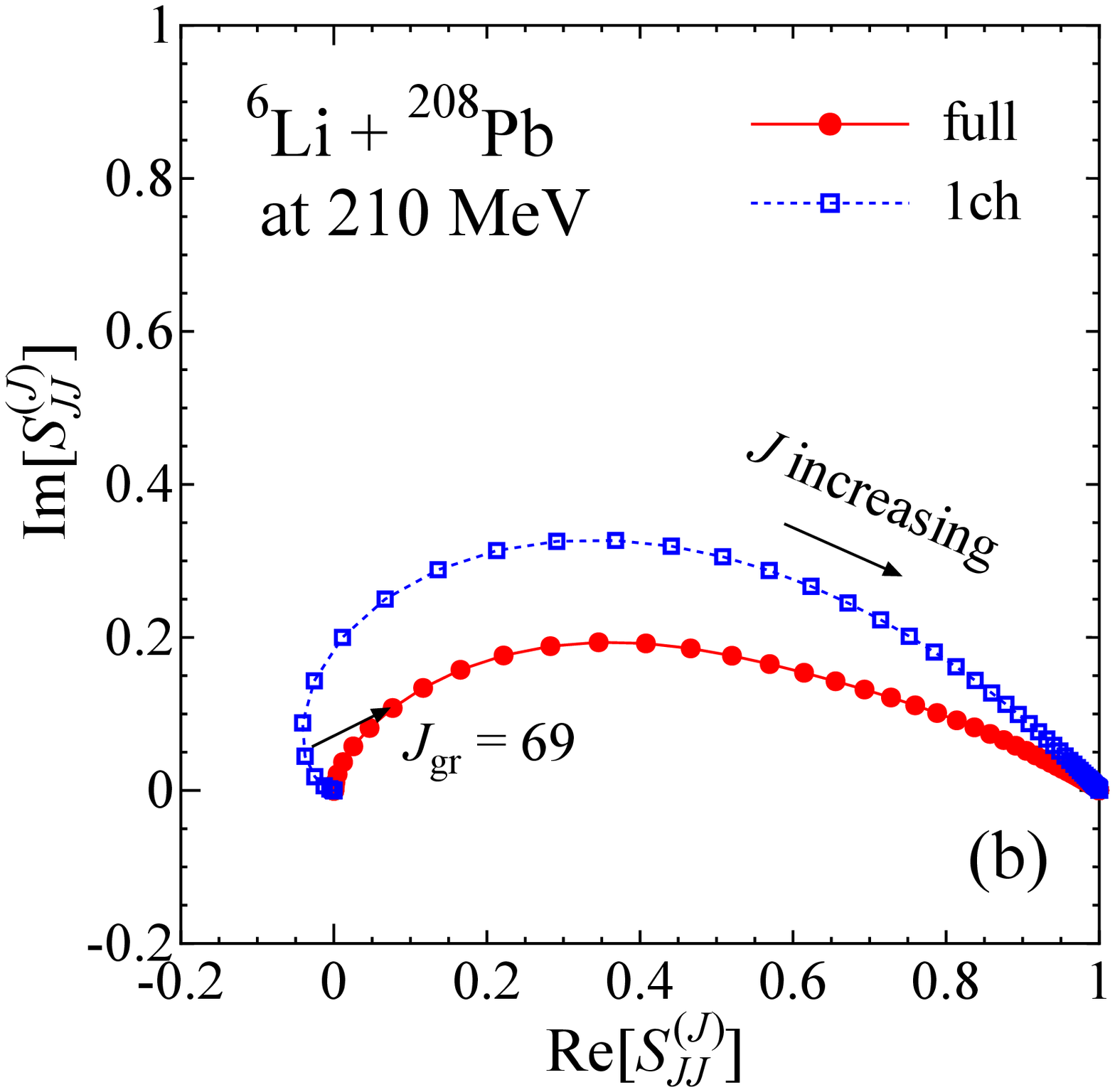}
 \caption{(Color online)
Elastic $S$-matrix elements for $^6$Li + $^{208}$Pb scattering (a) at 
$E_{\rm in}=39$ MeV and (b) at $E_{\rm in}=210$ MeV. 
 The open circles stand for the results of full CDCC calculations, 
 and the closed squares correspond to the results of 1ch calculations.
}
 \label{S_6Li+208Pb}
\end{center}
\end{figure}

We discuss the elastic $S$-matrix elements in Fig.~\ref{S_6Li+208Pb} 
for $^6$Li + $^{208}$Pb scattering at $E_{\rm in}=39$ and 210 MeV.
The elements are represented by $S_{L'L}^{(J)}$, where 
$J$  is the total angular momentum and $L$ $(L')$ is 
the initial (final) angular momentum regarding ${\Ve R}$ satisfying 
the relations $|J-1|\le L\le J+1$ and $|J-1|\le L'\le J+1$. 
Only the diagonal elements $S_{JJ}^{(J)}$ are plotted as a function of $J$ 
in Fig.~\ref{S_6Li+208Pb}. 
 Closed circles connected with solid lines 
 (open squares connected with dotted lines) stand for
 the results of full-CDCC (1ch) calculations. 
 Projectile breakup effects become small as $E_{\rm in}$ increases from 39 MeV 
 to 210 MeV, but the effects are still not negligible 
 at $E_{\rm in}=210$ MeV, as shown in Fig.~\ref{fig:el_208Pb_log_q}. 
 Projectile-breakup effects rotate $S_{JJ}^{(J)}$ clockwise 
 at the grazing total angular momentum $J_\mathrm{gr}$; note 
 that $J_\mathrm{gr}=17$ for 39 MeV and 69 for 210 MeV.
 The effects thus provide repulsive corrections to the results 
 of 1ch calculations, i.e., the folding potential.  
 This result is consistent with that of Ref. \cite{CDCC-review1} based on 
 three-body CDCC.

\subsection{Four- and three-body dynamics}
\label{Four-body dynamics}

In general, the ground and breakup states of $^{6}$Li consist of 
$d \alpha$ (two-body) and $np \alpha$ (three-body) configurations, 
although the main component of the ground state is a $d \alpha$ one. 
It is found in our previous paper \cite{Wat12} that 
$d$ little breaks up in $^6$Li-breakup processes of 
$^{6}$Li+$^{209}$Bi elastic scattering 
at $E_{\rm in} \approx E_{\rm b}^{\rm Coul} (\approx 30$ MeV). 
In this subsection, we confirm that the $d$-breakup suppression 
($d \alpha$ dominance) in $^6$Li elastic scattering 
is realized also for $ E_{\rm b}^{\rm Coul} < E_{\rm in} \la 210$ MeV. 
When $^6$Li elastic scattering is compared with $d$ one 
at the same incident energy per nucleon (the same incident velocity),  
$d$ breakup is strongly suppressed 
in the former scattering but not in the latter one. 
Note that in $d$ scattering 
$d$ breakup is significant at incident energies up to 
700 MeV \cite{CDCC-review1}.

The model space $P$ of CDCC calculations 
can be decomposed into the ground-state part $P_0$
and the breakup-state part $P^{*}$ as 
$P=P_{0}+P^{*}$ for 
\begin{align}
P_0=\ket{\Phi_0}\bra{\Phi_0}, \quad 
P^{*}=\sum_{\gamma=1}^N\ket{\Phi_\gamma}\bra{\Phi_\gamma}.
\end{align} 
For later discussion, $P^*$ is further divided into 
a subspace $P_{np\alpha}$ dominated by 
${np\alpha}$ configurations and a subspace $P_{d\alpha}$ 
by $d\alpha$ configurations. 
The subspaces are defined as follows. 
The probability of $d\alpha$ configurations in the breakup state 
$\Phi_\gamma$ is obtained by the overlap between $\Phi_\gamma$ and 
the $d$ ground state $\phi^{(d)}$: 
\begin{equation}
\Gamma^{(d\alpha)}_\gamma=
\left|\braket{\phi^{(d)}({\Ve y})|\Phi_\gamma({\Ve y},{\Ve r})}\right|^2. 
\end{equation}
We then define a breakup state 
with $\Gamma^{(d\alpha)}_\gamma>0.5$ 
($\Gamma^{(d\alpha)}_\gamma \le 0.5$) 
as a $d\a$-dominant ($np\a$-dominant) state. 
The subspace $P_{d\alpha}$ ($P_{np\alpha}$) is 
a model space spanned by $d\a$-dominant ($np\a$-dominant) breakup 
states. Consequently, the model space $P$ of CDCC calculations is 
expressed as 
\begin{equation}
P=P_{0}+P_{d\alpha}+P_{np\alpha}. 
\end{equation}

In the present calculation, the $d\a$ probability $\Gamma^{(d\alpha)}_0$ 
for the $^{6}$Li ground state is $0.70$ and the corresponding 
spectroscopic overlap $(\Gamma^{(d\alpha)}_0)^{1/2}$ is 0.83. 
This result is consistent with the value 0.86 of other 
three-body model calculation~\cite{Kik11} and 
the experimental estimation $0.85\pm0.04$~\cite{Til02}. 
The $^{6}$Li ground state is thus one of $d\alpha$ dominant states. 
The numbers of $np\alpha$- and $d\alpha$-dominant states 
in the $P^{*}$ space are 140 and 15, respectively, 
i.e., $P_{np\alpha}$ is much larger than $P_{d\alpha}$. 
This comes from the fact that the three-body phase space is 
larger than the two-body one.

\begin{figure}[htbp]
\includegraphics[width=0.8\figwidth,clip]{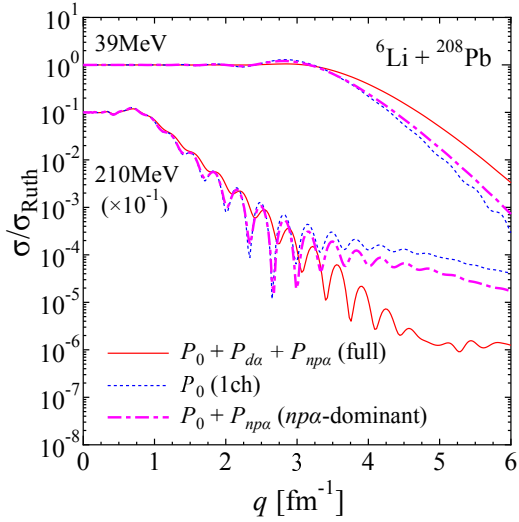}
\caption{(Color online) Elastic cross sections (normalized by the Rutherford cross section)
 for $^6$Li + $^{208}$Pb scattering at $E_{\rm in}=39$ and 210 MeV.
 The solid line represents the result of full four-body CDCC calculations, 
 whereas the dotted line denotes the results of 1ch calculations with 
 no breakup channel. 
 In the dot-dashed line, the model space $P_{d\alpha}$ is switched 
 off from the solid line.
\label{fig:el_208Pb_npA}}
\end{figure}

\begin{figure}[htbp]
\includegraphics[width=0.8\figwidth,clip]{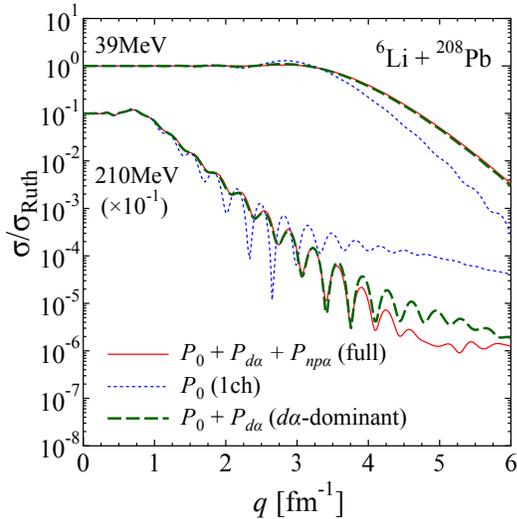}
\caption{(Color online) Same as Fig~\ref{fig:el_208Pb_npA}, 
 but in the dot-dashed line 
 the model space $P_{np\alpha}$ is switched 
 off from the solid line. 
\label{fig:el_208Pb_dA}}
\end{figure}

Now we confirm that the $d$-breakup suppression is 
realized also for $^{6}$Li scattering at $E_{\rm in} > E_{\rm b}^{\rm Coul}$. 
Figure~\ref{fig:el_208Pb_npA} shows differential cross sections for 
$^6$Li + $^{208}$Pb scattering at $E_{\rm in}=39$ and 210 MeV. 
The solid and dotted lines are the same as in Fig.~\ref{fig:el_208Pb_log_q}. 
When $d\a$-dominant states are switched off from full-CDCC calculations 
(solid line), we get the dot-dashed line. The line is close to the result of 
1ch calculations (dotted line) for each of $E_{\rm in}=39$ and 210 MeV. 
Figure~\ref{fig:el_208Pb_dA} is the same
as Fig.~\ref{fig:el_208Pb_npA}, but $np\a$-dominant states are 
switched off from full-CDCC calculations (solid line). 
The result (dot-dashed line) is close to the result of 
full-CDCC calculations (solid line). 
The $d\a$ dominance ($d$-breakup suppression) in $^{6}$Li breakup 
is thus confirmed.

\begin{figure}[htbp]
\includegraphics[width=0.95\figwidth,clip]{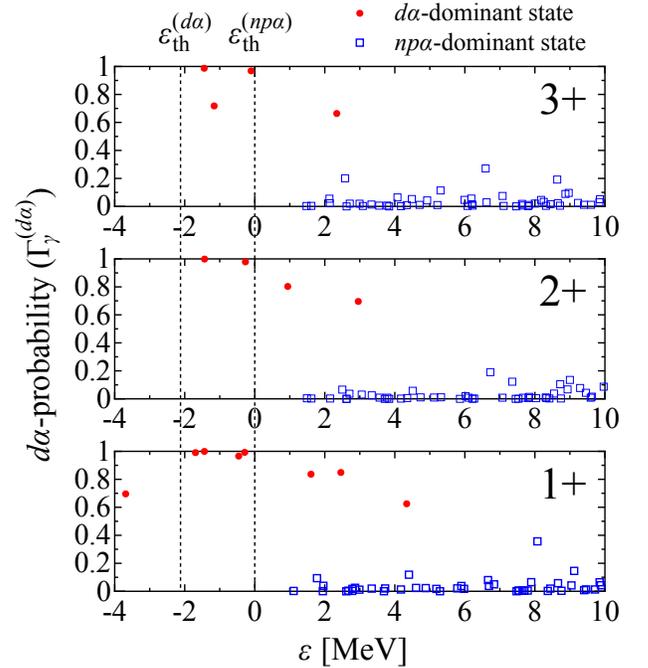}
\caption{(Color online) Distribution of $d\alpha$-probability in excitation spectrum.
The closed circles (open squares) correspond to
 the $d\alpha$-dominant states ($np\alpha$-dominant states); see text for details.
The threshold energies of $\varepsilon_\mathrm{th}^{(d\alpha)}=-2.2$ MeV and
$\varepsilon_\mathrm{th}^{(np\alpha)}=0$ MeV are also shown as the dotted line for reference.
\label{fig:over}}
\end{figure}

In order to understand the nature of the $d\a$ dominance, 
we plot the $\Gamma^{(d\alpha)}_\gamma$ 
as a function of $\varepsilon$ in Fig. \ref{fig:over}. 
In the region $\varepsilon_\mathrm{th}^{(d\alpha)} < \varepsilon 
< \varepsilon_\mathrm{th}^{(np\alpha)}$, the $\Gamma^{(d\alpha)}_\gamma$ are 
even larger than $\Gamma^{(d\alpha)}_0=0.70$. 
The $d\a$ dominance is thus somewhat developed in this region. 
Above $\varepsilon_\mathrm{th}^{(np\alpha)}$, some low-lying states keep 
$P^{(d\alpha)}_\gamma$ large; in fact, the values are comparable to 
$\Gamma^{(d\alpha)}_0$. Eventually, some of 15 $d\a$-dominant breakup states 
are concentrated on the low-lying part of excitation spectrum near 
$\varepsilon_\mathrm{th}^{(np\alpha)}$, 
whereas 140 $np\a$-dominant breakup states spread out in the spectrum. 

The $d\a$-dominant breakup states are thus
located in the lowlying part of excitation spectrum,
because the only $n+p$ subsystem has a bound state. 
Therefore, the low-lying $d\a$-dominant breakup states can work as a collective mode, 
since the breakup states have structures similar to 
the $^{6}$Li ground state and the transitions between them become strong. 
In fact, the coupling potentials $U_{\gamma0}$ from  the ground state 
to the low-lying $d\a$-dominant states are much larger than 
the $U_{\gamma0}$ to the $np\a$-dominant breakup states. 
This property means that the incident flux in the elastic channel 
mainly goes to 
the low-lying $d\a$-dominant breakup channels and comes back to 
the elastic channel. 
This is the reason why the $d\a$ dominance ($d$-breakup suppression) is 
realized in $^{6}$Li breakup independently of $E_{\rm in}$.

\section{summary}
\label{summary}

Four-body ($n+p+\alpha+ \mathrm{target}$) dynamics 
of $^6$Li elastic scattering was analyzed over a wide range of $E_{\rm in}$ 
with four-body CDCC. Four-body CDCC well reproduced 
measured elastic and total reaction 
cross sections with no adjustable parameter 
for $^{6}$Li+$^{209}$Bi scattering 
at $E_{\rm in}=24$--50 MeV and $^{6}$Li+$^{208}$Pb scattering 
at $E_{\rm in}=29$--210 MeV. 
In the wide $E_{\rm in}$ range, $^6$Li breakup is significant and 
provides repulsive corrections to the folding potential. 
The problem addressed by Satchler and Love 
is thus clearly solved by the present four-body CDCC calculation.

In our previous work \cite{Wat12}, we found that $d$ breakup is 
strongly suppressed in $^{6}$Li scattering near 
the Coulomb-barrier energy. 
In the present paper, we confirmed that the $d$-breakup suppression is 
realized for any $E_{\rm in}$, as far as 
$^{6}$Li breakup itself is significant. This mechanism can be 
understood as follows. 
Some of $d\a$-dominant breakup states 
are concentrated on the low-lying part of excitation spectrum and work 
as a collective-excitation mode effectively. 
In fact, $^{6}$Li breakup is mainly induced 
by strong transitions to the breakup states, and consequently, 
$d$ breakup is suppressed in $^{6}$Li breakup processes.

The $d$-breakup suppression may indicate that 
$^{6}$Li scattering is described effectively by the $d+\a+T$ three-body 
model, if the following two points are satisfied. 
As for the potential between $d$ and $T$, 
we should use the single-folding potential 
obtained by folding $U_n$ and $U_p$ with the $d$ ground state, 
since the folding potential does not include $d$-breakup effects. 
As for the projectile ($^{6}$Li) radius important for 
elastic scattering, the value calculated with the $d+\a$ two-body model 
should be consistent with that with the $n+p+\a$ three-body model. 
A successful example is shown in our previous paper \cite{Wat12}. 
Further analyses along this line are quite interesting.

\section*{Acknowledgements}
The authors are grateful to K. Minomo and A. M. Moro for fruitful discussions.
This work was supported by JSPS KAKENHI Grant Numbers
 25$\cdot$4319, 25400255, 26400278.

\appendix

\section{Coulomb-breakup effects}
\label{sec:CoulombBU}

\begin{figure}[htbp]
\includegraphics[width=0.8\figwidth,clip]{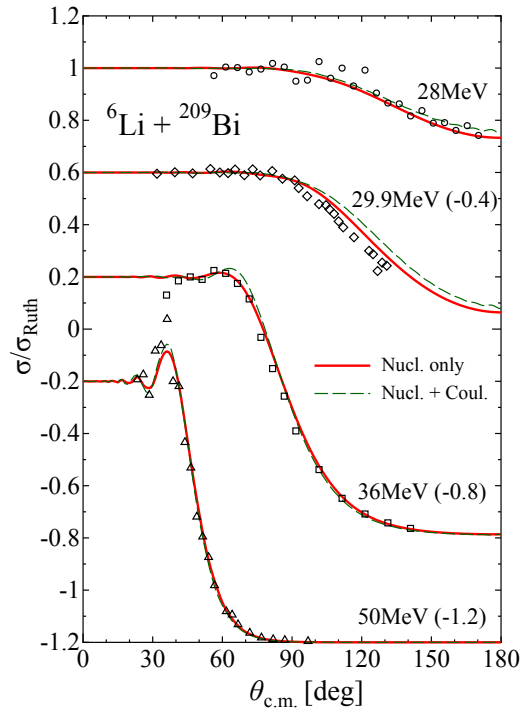}
\caption{(Color online) Coulomb-breakup effects 
 on $^6$Li + $^{209}$Bi elastic cross section at 28--50 MeV.
 The solid and dashed lines show the results of CDCC calculations 
 without and with Coulomb breakup.
 The symbols are experimental data taken from Refs.~\cite{Agu00,Agu01,San11}.
\label{fig:el_Coul}}
\end{figure}

Here, we check Coulomb-breakup effects on 
$^6$Li + $^{209}$Bi elastic scattering. 
In Eq.~(\ref{eq:H4-ver2}), the Coulomb interaction  is then replaced back to 
\begin{equation}
\frac{e^2Z_{\rm Li}Z_{\rm T}}{R}\rightarrow
\frac{e^2Z_p Z_{\rm T}}{R_p}+\frac{e^2Z_\alpha Z_{\rm T}}{R_\alpha}.
\end{equation}
Figure~\ref{fig:el_Coul} shows Coulomb-breakup effects 
on differential cross sections 
for $^6$Li + $^{209}$Bi elastic scattering at $E_{\rm in}=28$--50 MeV. 
The solid and dashed lines correspond to CDCC calculations 
without and with Coulomb breakup, respectively. 
The difference between the two lines is tiny, indicating that 
Coulomb-breakup effects are quite small.
This comes from the lack of electric dipole transitions; 
see ~Appendix~\ref{sec:E1} for the theoretical discussion.
Coulomb breakup effects are thus suppressed in $^6$Li elastic scattering 
compared with $^6$He scattering~\cite{Mat06,Kee03}.

\section{Electric dipole transitions in the $N+N+\alpha$ three-cluster model}
\label{sec:E1}

\begin{figure}[htbp]
\includegraphics[width=0.4\figwidth,clip]{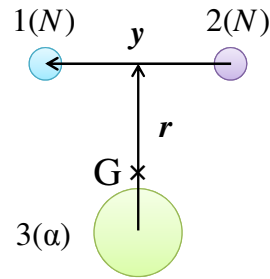}
\caption{(Color online) 
The $N+N+\alpha$ three-cluster model and its coordinates.
\label{fig:coordinate_3b}}
\end{figure}

Coulomb breakup is mainly induced by the electric dipole transition. 
It is well known that the transition strength vanishes 
in the $d+\a$ model for $^6$Li, since the mass ratio $m_d/m_\a$ equals to 
the charge one $Z_d/Z_\a$~\cite{Kee03,Buc77}. 
The vanishment is true also for the $n+p+\a$ model, as shown later. Let us 
consider the $N+N+\alpha$ three-cluster model illustrated in Fig. 
\ref{fig:coordinate_3b} in which two nucleons are labeled by 
1 and 2 and $^{4}$He is by 3. In the model, the dipole operator ($D_\mu$) 
is given by
\begin{equation}
D_\mu=\sum_{i=1}^{2}(1/2-\tau_{iz})ex_iY_{1\mu}(\hat{\Ve x}_i)+2ex_3Y_{1\mu}(\hat{\Ve x}_3),\label{eq:Dipole}
\end{equation}
where $\tau_{iz}=1/2$ ($-1/2$) for $N=n$ ($p$), and 
${\Ve x}_i$ is the coordinate of cluster $i$ from the center of mass (G): 
\begin{align}
{\Ve x}_1&=\frac{2}{3}{\Ve r}+\frac{1}{2}{\Ve y},\\
{\Ve x}_2&=\frac{2}{3}{\Ve r}-\frac{1}{2}{\Ve y},\\
{\Ve x}_3&=-\frac{1}{3}{\Ve r}.
\end{align}
The total isospin $T$ and its $z$ component $T_z$  
of $^6$Li are zero, and the isospin component of the $^6$Li ground state 
is described by 
\begin{equation}
\ket{T T_z}=\ket{00}=\frac{1}{\sqrt{2}}(\ket{np}+\ket{pn}) .
\end{equation}
The expectation value of $D_\mu$ for $\ket{00}$ is then 
\begin{equation}
\braket{00|D_\mu|00}=\frac{e}{2}\sum_{i=1}^{2}x_iY_{1\mu}(\hat{\Ve x}_i)
+2ex_3Y_{1\mu}(\hat{\Ve x}_3)=0, 
\end{equation}
since 
\begin{align}
x_1Y_{1\mu}(\hat{\Ve x}_1)
&=\frac{2}{3}rY_{1\mu}(\hat{\Ve r})+\frac{1}{2}yY_{1\mu}(\hat{\Ve y}),\label{eq:E1_1}\\
x_2Y_{1\mu}(\hat{\Ve x}_2)
&=\frac{2}{3}rY_{1\mu}(\hat{\Ve r})-\frac{1}{2}yY_{1\mu}(\hat{\Ve y}),\label{eq:E1_2}\\
x_3Y_{1\mu}(\hat{\Ve x}_3)
&=-\frac{1}{3}rY_{1\mu}(\hat{\Ve r}).
\label{eq:E1_3}
\end{align}

\end{document}